\documentstyle[twocolumn,aps,graphics,epsfig,prb]{revtex}
\begin{document}

\tighten
\draft
\twocolumn[
\hsize\textwidth\columnwidth\hsize\csname @twocolumnfalse\endcsname

\title{Dynamical mean-field theory of a double-exchange model with
diagonal disorder}
\author{B.M. Letfulov$^*$ and J.K. Freericks$^\dagger$}
\address{$^*$Institute of Metal Physics, Kovalevskaya Str. 18,
Yekaterinburg, 620219, Russia}
\address{$^\dagger$Department of Physics, Georgetown University, Washington, DC
20057, U.S.A.}
\date{\today}
\maketitle

\widetext
\begin{abstract}
We present a simplified model for the colossal magnetoresistance in doped
manganites by exactly solving a double-exchange model (with Ising-like local
spins) and quenched binary disorder within dynamical mean field theory.  We
examine the magnetic properties and the electrical and thermal transport. Our
solution illustrates three different physical regimes: (i) a weak-disorder
regime, where the system acts like a renormalized double-exchange system
(which is insufficient to describe the behavior in the manganites); (ii)
a strong-disorder regime, where the system is described by strong-coupling
physics about an insulating phase (which is the most favorable for
large magnetoresistance); and (iii) a transition region of moderate
disorder, where both double-exchange and strong-coupling effects are
important.  We use the thermopower as a stringent test for the applicability
of this model to the manganites and find that the model is unable to properly
account for the sign change of the thermopower seen in experiment.
\end{abstract}
]
\narrowtext

\section{Introduction}
\label{Intro}
An enormous interest in the family of doped manganese oxides
La$_{1-x}$A$_x$MnO$_3$ (in which A stands for Ca, Sr or Pb) has been created
by the colossal magnetoresistance (CMR)
exhibited in samples with doping levels in the range\cite{r1}
$0.15<x<0.4$. In such a doping region,
there is a characteristic temperature $T_p$ where the resistivity $\rho$
has a peak; the CMR materials display metallic behavior (defined by
$d\rho/dT>0$) for $T<T_p$, and insulating behavior (defined by $d\rho/dT
<0$) for $T>T_p$ (except for LSMO at $x\approx 0.3$).  Hence there is a
metal-insulator transition (MIT)  or crossover at $T_p$. When placed in an
external magnetic field $H$, the resistivity
is strongly suppressed and the temperature at which the resistivity has
a peak ($T_p$) increases.  The magnitude of the relative magnetoresistance can
be extremely large and can attain 99\% or more in some samples.

It is widely accepted that the transport properties of these systems are closely
related to their magnetic properties. The temperature of the MIT is
close to the Curie temperature $T_c$ so one can say that these materials
are metallic in the ferromagnetic phase and are insulating in the paramagnetic
phase. The itinerant-electron and the local-spin states are correlated
by the double-exchange mechanism~\cite{r2,r3,r4,r5}
(which is one type of indirect exchange interaction between local
spins via itinerant electrons). Double exchange
consists of a cooperative effect where the motion of an
itinerant electron favors the formation of ferromagnetic order of the
local spins and, vice versa, the presence of ferromagnetic order
facilitates the motion of the
itinerant electrons. Hence, the resistivity of a double-exchange system
will increase when the temperature is increased through the Curie point.
This is in qualitative agreement with the experimental data on the manganites.

Unfortunately, it is also well known  that double exchange
alone cannot explain the quantitative features of the
temperature dependence of the resistivity through the
entire temperature range observed in the manganites~\cite{r6} (especially
near the MIT). There are two proposed theoretical resolutions of this problem.
The first one is that a large Jahn-Teller lattice distortion
is responsible for the anomalous transport properties~\cite{r6,r7,r8,r9,r10}.
The lattice distortion causes a metal-insulator transition via a strong
(exponential) polaronic narrowing of the conduction-electron band. This
polaronic narrowing also leads to a decrease of the Curie temperature because
the double-exchange mechanism for ferromagnetic order is reduced when the
bandwidth of the conduction electrons is narrowed.

The polaronic mechanism has been criticized by a number of authors.
Varma~\cite{r11} noticed that there exist
double-exchange systems, such as TmSe$_x$Te$_{1-x}$, in which the transport
anomalies seen in the manganites also occur, but a Jahn-Teller distortion is
forbidden by symmetry. Furukawa~\cite{r12} showed that a small-polaron
picture leads to a strong suppression of the ferromagnetic transition
temperature and the estimate of $T_c$ due to double exchange was
incorrectly calculated in Ref.~\onlinecite{r6}
(Ref.~\onlinecite{r13} reaches a similar conclusion).
Further difficulties arise from the fact that LSMO does not have a
metal-insulator transition\cite{r14} at $x\sim0.3$.
Furukawa claims that LSMO [which has a
relatively high value for $T_c$ (about $380^o$K)] is a canonical double-exchange
system (note that LSMO does show\cite{r14} a doping crossover from
metallic behavior at $x\sim0.3$ to insulating behavior at
$x\sim0.15$). Finally, an analysis\cite{r15,r16} of the
longitudinal and Hall resistivities in
LCMO and LPMO cannot be explained in the small-polaron picture for
the temperature range near $T_p$ or for high temperatures ($T\gg T_p$).

In the second proposed
resolution\cite{r11,r16,r17,r18,r19,r20,r21,r22,r23}, the insulating
behavior is caused by a combination of both magnetic disorder (due to the lack
of ferromagnetic alignment in the paramagnetic phase) and
nonmagnetic ionic disorder (due to the doping of the ``A'' metal).
The magnetic disorder arises from the ``random'' double-exchange factor
$\cos\Theta/2$ in the electronic hopping (where $\Theta$ is the angle between
local spins). This is an
off-diagonal disorder that can lead to a Lifshitz localization~\cite{r24}
of the charge carriers. Sheng et al.\cite{r18,r19} have found that this
off-diagonal disorder is insufficient to localize the electronic states
at the Fermi level for moderate doping (which agrees with the claim that
the double-exchange mechanism alone cannot describe the behavior of the
manganites). The nonmagnetic disorder comes
from the ionic doping of the A$^{2+}$ ions (i.e. from randomness at the
chemical substitution of
La by A$^{2+}$) which leads to a ``random'' local potential for the
charge carriers. This substitutional disorder is always present in the
doped materials,
and it is physically meaningful to speak only about
substitutional-disorder-averaged quantities\cite{r25}.
This ionic disorder is a diagonal disorder that
can lead to an Anderson localization of the
charge carriers. One-parameter scaling
calculations\cite{r18} show that in the presence of a suitable strength
of the ionic disorder, the magnetic disorder will cause the localization
of electrons at the Fermi surface and induce a metal-insulator transition near
$T_c$. (However, there is experimental evidence\cite{r26} that
Anderson localization is not the cause of the metal-insulator transition
in  La$_{0.67}$Ca$_{0.33}$MnO$_3$).

Zhong et al.\cite{r21} have used dynamical mean field theory to study the
metal-insulator transition in the manganites in the framework of an $s-d$ model
with classical local spins and doping-induced disorder (see also
Ref.~\onlinecite{r23}). They were able to show that a MIT is possible
for a binary-alloy distribution of the ionic energy levels due to a splitting
of the electronic band (correlated by the double-exchange process)
into completely filled and empty subbands
at some critical value of disorder strength.
Such a distribution was also used\cite{r25} to study the Hubbard
model with substitutional disorder within dynamical mean field theory.

In this contribution, we consider the magnetic and transport properties
of a simple double-exchange system with diagonal disorder.
The system is described by the following Hamiltonian
\begin{equation}
	{\cal H}=\sum_{i\sigma}\epsilon_ic^{\dag}_{i\sigma}c^{}_{i\sigma}+
	\sum_{ij\sigma}t_{ij}c^{\dag}_{i\sigma}c^{}_{j\sigma},
	\label{e1}
\end{equation}
where the $c$-operators are composite operators
\begin{equation}
	c_{i\sigma}=\frac{1}{2}(1+\sigma s^z_i)a_{i\sigma},
	\label{e2}
\end{equation}
with $s^z_i$ the z-component of the local spin ($S=1/2$)
described by the diagonal Pauli matrix,
$(s^z_i)^2=1$, and $a^{}_{i\sigma}(a^{\dag}_{i\sigma})$ the ordinary Fermi
annihilation (creation) operator for an itinerant electron with spin
projection $\sigma$ at lattice site $i$. The first term in Eq.~(\ref{e1})
describes the doping-induced (diagonal) disorder (the energies $\epsilon_i$
are chosen from a disorder distribution) and the second term represents a
simplified version of the quantum-mechanical double-exchange mechanism for
ferromagnetic ordering of local
spins. This term can be obtained from an $s-d$ model with local spins
that are described by Ising spins in the limit of infinitely strong
exchange interaction between itinerant and localized electrons.
This simplification of describing the local spins by
Ising variables conserves the main feature of double exchange: the second term
in Eq.~(\ref{e1}) only allows the transfer of itinerant electrons with spin
parallel to the local spin at every site of the lattice
[see Eq.~(\ref{e2})].  But, these simplified operators do
not allow any spin-flip processes. Such processes can be important
at low temperatures where the thermodynamics of the system is governed
by spin-wave excitations. However, since dynamical mean-field theory
cannot describe spin waves, including such effects is beyond the scope of
this work.  Spatial correlations between the local spins of the form
$\langle s^z_is^z_j\rangle$ [which is contained in Eq.~(\ref{e1})] should be
important near the Curie point\cite{r6,r27} $T_c$, but they are also beyond
the capabilities of dynamical mean field theory.

This paper is organized as follows: The dynamical mean-field theory
equations for the system
are presented in Section II. The binary probability distribution
for the doping-induced disorder and simplifications for this disorder
averaging are discussed in Section III.
In Section IV, the influence of disorder on the magnetic properties (decreases
of $T_c$, the paramagnetic susceptibility, and the magnetization of the
local-spin
subsystem) are investigated. Here, we show that for strong disorder, the
double-exchange mechanism of ferromagnetic ordering is replaced by a
disorder-induced ferromagnetism and we describe three characteristic disorder
regimes: (i) the weak-disorder regime where the double-exchange mechanism
dominates; (ii) the strong-disorder regime where the
magnetic and transport properties are determined by strong-coupling physics
from an insulating phase; and (iii) a transition
regime where both mechanisms are important. The main transport properties
(resistivity, thermopower, and thermal conductivity) in these three regimes
(along with the magnetoresistivity) are discussed in Sections V and VI,
respectively. Section VII contains our concluding remarks.

\section{Formalism for the dynamical mean field theory}
\label{basic}
The dynamical mean field theory
equations for the system described by Eq.~(\ref{e1}) can be obtain in two ways:
(i) a diagrammatic technique for $c$-operators~\cite{r28} can be
combined with disorder averaging\cite{r29} in the
limit $z\to\infty$ ($z$ is the coordination number), or (ii) one can work
directly
with a local effective action $S_{eff}$. The first method yields a direct
calculation of the dynamical mean field equations for the disorder-averaged
band Green's function and for the
magnetization of the local-spin subsystem by exactly summing all
nonvanishing graphs as $z\to\infty$. This procedure is cumbersome, so we
will use the effective action approach here.

Since the anticommutator of $c$-operators is not a number, it is
convenient to begin with the original $s-d$ model with Ising spins and
diagonal disorder:
\begin{eqnarray}
	{\cal H}&=&-\frac{1}{2}h\sum_{i}s^z_i+\sum_{i\sigma}(\epsilon_i-\mu-
	\frac{1}{2}\sigma H)a^{\dag}_{i\sigma}a^{}_{i\sigma}\cr
        &+&\sum_{ij\sigma}
        t_{ij}a^{\dag}_{i\sigma}a^{}_{j\sigma}-\frac{1}{2}I\sum_{i\sigma}
        s^z_ia^{\dag}_{i\sigma}\sigma a^{}_{i\sigma},
	\label{e3}
\end{eqnarray}
where we have introduced two external magnetic fields ($h$ acts on the
local-spin subsystem and $H$ acts on the itinerant-electron subsystem).
This is done to allow derivatives with respect to the fields to be
calculated properly.  In the end results, we set the two fields equal to
the true external magnetic field.  We will take the limit where the $s-d$
exchange parameter $I$ becomes infinitely large $(I\rightarrow\infty$),
because this is the regime where the $s-d$ Hamiltonian is mapped onto the
double-exchange Hamiltonian.

The mathematical structure of the $s-d$ Hamiltonian in
Eq.~(\ref{e3}) is similar to that of the
Falicov-Kimball model\cite{falicov-kimball}
which can be solved exactly in infinite dimensions~\cite{r30}.
The procedure is to first solve the atomic problem in an
external time-dependent field and then adjust the field so that the
atomic Green's function equals the local lattice Green's function.  The local
effective action for this atomic problem is
\begin{eqnarray}
        S_{eff}(\epsilon)&=&-\frac{1}{2}\beta hs^z+\sum_{\sigma}\int
        \limits_{0}^{\beta}d\tau a^{\dag}_{\sigma}(\tau)\biggl(\epsilon-\mu-
        \frac{1}{2}\sigma H\cr
        &-&\frac{1}{2}I\sigma s^z\biggr)a^{}_{\sigma}(\tau)
        +\sum_{\sigma}
        \int\limits_{0}^{\beta}d\tau\int\limits_{0}^{\beta}d\tau'a^{\dag}
        _{\sigma}(\tau)\cr
        &\times&\biggl(\frac{\partial}{\partial\tau}\delta(\tau-\tau')+
	\Lambda_{\sigma}(\tau-\tau')\biggr)a^{}_{\sigma}(\tau'),
	\label{e4}
\end{eqnarray}
with $\beta=1/T$ and $\Lambda_\sigma(\tau)$ is the time-dependent field.
The partition function becomes
\begin{equation}
	Z_{eff}(\epsilon)=\mbox{Tr}\int Da^{\dag}_{\sigma}Da^{}_{\sigma}
	e^{-S_{eff}(\epsilon)},
	\label{e5}
\end{equation}
where the trace is taken over $s^z$.  The disorder-averaged free energy becomes
\begin{equation}
	F_{eff}=-T\langle\ln Z_{eff}(\epsilon)\rangle\equiv-T\int d\epsilon
	P(\epsilon)\ln Z_{eff}(\epsilon).
	\label{e6}
\end{equation}
Here, $P(\epsilon)$ is a probability distribution function for the (random)
atomic energies $\epsilon_i$ and the angle brackets $\langle\ldots\rangle$
denote the disorder averaging.

The disorder-averaged local band Green's function $\langle{\cal G}_{n\sigma}
\rangle$
is determined by a functional derivative with respect to the atomic field.
In Fourier space, we have
\begin{equation}
	\langle{\cal G}_{n\sigma}\rangle=\beta\frac{\partial F_{eff}}
	{\partial\Lambda_{n\sigma}},
	\label{e7}
\end{equation}
where $\Lambda_{n\sigma}\equiv\Lambda_{\sigma}(i\omega_n)$ is the Fourier
transform of $\Lambda_{\sigma}(\tau)$ and $\omega_n=(2n+1)\pi T$
is the Fermionic Matsubara frequency.

In taking the limit $I\to\infty$ in Eq.~(\ref{e5}),
we must first renormalize the chemical potential
($\mu+I/2\to\mu$). Then Eq.~(\ref{e7}) becomes
\begin{equation}
	\langle{\cal G}_{n\sigma}\rangle=\int d\epsilon P(\epsilon)\frac{
	\frac{1}{2}[1+\sigma m(\epsilon)]}{a_{n\sigma}+\frac{1}{2}\sigma H-
	\epsilon},
	\label{e8}
\end{equation}
where
\begin{equation}
	a_{n\sigma}=i\omega_n+\mu-\Lambda_{n\sigma},
	\label{e9}
\end{equation}
is the inverse of the effective medium and
\begin{equation}
	m(\epsilon)=\tanh\frac{1}{2}\bigl[\beta h+\beta H/2+
	\lambda_F(\epsilon)\bigr],
	\label{e10}
\end{equation}
is the magnetization of the local spins (when the band electrons
have disorder energy $\epsilon$) and
\begin{equation}
        \lambda_F(\epsilon)=\sum_{n}\ln\frac{a_{n\uparrow}+H/2-\epsilon}
	{a_{n\downarrow}-H/2-\epsilon}.
	\label{e11}
\end{equation}
The total magnetization is defined to be
\begin{equation}
	m=\langle s^z_i\rangle=-\frac{\partial F_{eff}}{\partial(h/2)}=
	\int d\epsilon P(\epsilon)m(\epsilon).
	\label{e12}
\end{equation}

Note that the expression in Eq.~(\ref{e10}) for the magnetization contains a
hyperbolic tangent, just like in the mean-field theory of an Ising ferromagnet.
Therefore, $\lambda_F(\epsilon)$ can be identified as
an internal molecular field acting on the
local spins. Eqs.~(\ref{e11}) and (\ref{e8}) show that this molecular
field is completely determined by the properties of the itinerant-electron
subsystem. Hence, the magnetic
and transport properties of the double-exchange system are correlated.

In infinite dimensions, the inverse of the effective medium $a_{n\sigma}$,
local self energy $\Sigma_{n\sigma}=\Sigma_{\sigma}(i\omega_n)$,
and the local Green's function $\langle{\cal G}_{n\sigma}\rangle$ are
related by
\begin{equation}
	\langle{\cal G}_{n\sigma}\rangle^{-1}=a_{n\sigma}-\Sigma_{n\sigma}.
	\label{e13}
\end{equation}
The self consistency relation equates the atomic Green's function with the
local Green's function of the lattice.  The latter can be calculated from
the local self energy by summing over all momentum.  Since the self
energy is independent of momentum, we find
\begin{equation}
        \langle{\cal G}_{n\sigma}\rangle=\frac{1}{N}\sum_k
        {\cal G}_{\sigma}(k,i\omega_n)
        =\int\limits_{-\infty}^{\infty}
        \frac{dxD^0(x)}{i\omega_n+\mu-\Sigma_{n\sigma}-x},
	\label{e14}
\end{equation}
where $D^0(x)$ is the density of states of the noninteracting itinerant
electrons. We choose to examine the infinite-coordination Bethe lattice, where
\begin{equation}
        D^0(x)=\frac{1}{2\pi t^{*2}}\sqrt{(2t^*)^2-x^2},
	\label{e15}
\end{equation}
with $2t^*=1$ chosen to be the energy unit.  The integral in Eq.~(\ref{e14})
can be computed exactly yielding
\begin{equation}
\langle{\cal G}_{n\sigma}\rangle=\frac{i\omega_n+\mu-\Sigma_{n\sigma}}{2t^{*2}}
\pm \frac{\sqrt{(i\omega_n+\mu-\Sigma_{n\sigma})^2-4t^{*2}}}{2t^{*2}}.
\label{integral}
\end{equation}
Replacing $\Sigma_{n\sigma}$ by $a_{n\sigma}-
\langle{\cal G}_{n\sigma}\rangle^{-1}$ and solving for the Green's function
gives
\begin{equation}
	t^{*2}\langle{\cal G}_{n\sigma}\rangle=i\omega_n+\mu-a_{n\sigma},
	\label{e16}
\end{equation}
and
\begin{equation}
	\Sigma_{n\sigma}=a_{n\sigma}-\frac{t^{*2}}{i\omega_n+\mu-a_{n\sigma}}.
	\label{e17}
\end{equation}
Substituting (\ref{e16}) into (\ref{e8}), we obtain the following final
equation for $a_{n\sigma}$:
\begin{equation}
	a_{n\sigma}=i\omega_n+\mu-t^{*2}\int\limits_{-\infty}^{\infty}
	d\epsilon P(\epsilon)\frac{\frac{1}{2}[1+\sigma m(\epsilon)]}
	{a_{n\sigma}+\frac{1}{2}\sigma H-\epsilon}.
	\label{e18}
\end{equation}
This equation is simpler than the corresponding system of
equations analyzed by Zhong et al.\cite{r21}. However, we believe that our
approach captures the main features of this system, as we described in
Section I.

The chemical potential $\mu$ is adjusted to give the correct filling for the
itinerant electrons:
\begin{eqnarray}
	n&=&-\frac{1}{\pi}\sum_\sigma\int_{-\infty}^{\infty}d\omega f(\omega)
        {\rm Im}\langle{\cal G}_\sigma(\omega-\mu)\rangle\cr
     &=&\frac{1}{\pi t^{*2}}\sum_{\sigma}\int\limits_{-\infty}^{\infty}
        d\omega f(\omega){\rm Im}a_{\sigma}(\omega-\mu),
	\label{e19}
\end{eqnarray}
where
\begin{equation}
	f(\omega)=\frac{1}{\exp{\beta(\omega-\mu)}+1},
	\label{e20}
\end{equation}
is the Fermi-Dirac function and $a_{\sigma}(\omega)$ is the solution of
Eq.~(\ref{e18}) (evaluated on the real-frequency axis).

Using Eq.~(\ref{e18}), we can also evaluate both the magnetization $m$ and the
paramagnetic susceptibility $\chi$ of the local-spin subsystem:
\begin{equation}
	\chi=\left.\frac{dm}{dh}\right|_{h=H=m=0}=\int d\epsilon P(\epsilon)
	\left.\frac{dm(\epsilon)}{dh}\right|_{h=H=m=0},
	\label{e21}
\end{equation}
where $dm(\epsilon)/dh$ is defined by the following integral equation
\begin{eqnarray}
        \frac{dm(\epsilon)}{dh}&=&\frac{1}{2}\beta-\frac{1}{2}t^{*2}
        \sum_{n}\frac{1}{a_n-\epsilon}\cr
        &\times&\frac{1}{1-{\cal A}_n}\int d\epsilon'
	\frac{P(\epsilon')}{a_n-\epsilon'}\frac{dm(\epsilon')}{dh},
	\label{e22}
\end{eqnarray}
with
\begin{equation}
	{\cal A}_n=\frac{1}{2}t^{*2}\int d\epsilon P(\epsilon)\frac{1}
	{(a_n-\epsilon)^2}.
	\label{e23}
\end{equation}
We note that an integral equation for a two-particle correlation
function is expected for disordered systems described by the
Falicov-Kimball model\cite{r32}.

\section{Binary probability distribution}
\label{distr}
An analysis of the electronic properties of the manganites
shows\cite{r21,r23} that a binary distribution for doping-induced
disorder can be approximately suitable for the doped materials.
This distribution is written in a symmetric form as\cite{r25}
\begin{equation}
	P(\epsilon_i)=(1-x)\delta(\epsilon_i+\frac{1}{2}\Delta)+
	x\delta(\epsilon_i-\frac{1}{2}\Delta),
	\label{e24}
\end{equation}
where $x$ is the fraction of the sites having an additional local potential
$\Delta$ (disorder strength) due to the ionic doping [the symmetric form
requires a renormalization of the chemical potential $\mu\to\mu-(1-x)\Delta/2$].
It is seen from
Eq.~(\ref{e24}) that this choice for the distribution behaves electronically
as a coherent superposition of its end-point compounds\cite{r21,r33}.

If the chemical substitution of the ions also causes the appearance of holes
in the band, then there is a correlation between the electron density and the
concentration of dopant sites
\begin{equation}
1-x=\frac{n}{\nu},
\end{equation}
where $\nu$ is the number of electronic states per lattice site. In the system
we consider, double occupation by itinerant electrons is excluded so $\nu=1$.
Therefore, we have $1-x=n$ in the probability distribution of Eq.~(\ref{e24}).
This constraint is important to allow a MIT.

The basic dynamical-mean-field equations are simplified for the binary
distribution. In particular, Eq.~(\ref{e18}) becomes
        $$
        a_{n\sigma}=i\omega_n+\mu-\frac{1}{2}t^{*2}\biggl\{\frac{(1-n)
	[1+\sigma m(\frac{1}{2}\Delta)]}{a_{n\sigma}+\frac{1}{2}\sigma H-
        \frac{1}{2}\Delta}
        $$
\begin{equation}
        +\frac{n[1+\sigma m(-\frac{1}{2}\Delta)]}
	{a_{n\sigma}+\frac{1}{2}\sigma H+\frac{1}{2}\Delta}\biggr\},
	\label{e25}
\end{equation}
where $m(\Delta/2)$ and $m(-\Delta/2)$ depend on the complete set
$\{ a_{n\sigma}\}$. Aside from the factors of $[1+\sigma m(\pm\Delta/2)]/2$,
this result is identical to that of an annealed binary alloy problem, as
first solved\cite{r30} by Brandt and Mielsch.

The main feature of the probability distribution (\ref{e24}) [with $1-x=n$]
is that the chemical potential is located in the
gap for any electron density (in the paramagnetic phase)
 when the conduction band is split by strong disorder
$\Delta>\Delta_c$. Indeed, when $\Delta=0$, the local band Green's function
has the form
\begin{equation}
	{\cal G}_{\sigma}(\omega)=\frac{1}{2t^{*2}}\bigl\{\omega+\mu-
	\sqrt{(\omega+\mu)^2-2t^{*2}(1+\sigma m)}\bigr\}.
	\label{e26}
\end{equation}
This shows that the bandwidth of the pure double-exchange system is
equal to $4t^*\sqrt{(1+m)/2}$ for the (majority)
spin-up electrons and $4t^*\sqrt{(1-m)/2}$
for the (minority) spin-down electrons. At $T=0$ (where $m=1$), there are no
spin-down electrons (the spin-down electron bandwidth is zero), and the
spin-up electrons act like free electrons with the
full value of $4t^*=W$ for the bandwidth. This ferromagnetic ordering promotes
the motion of the electrons. Increasing the temperature destroys the
ferromagnetism;
the bandwidth for the spin-up electrons is decreased and the bandwidth for
spin-down electrons is increased so that they are both equal to
$4t^*\sqrt{1/2}$ in the high-temperature paramagnetic phase\cite{r27}. In
the paramagnetic phase the bandwidth and density of states are
independent of temperature\cite{leinung}.

\begin{figure}[htb]
\begin{center}
        \resizebox{0.45\textwidth}{!}{%
        \includegraphics*{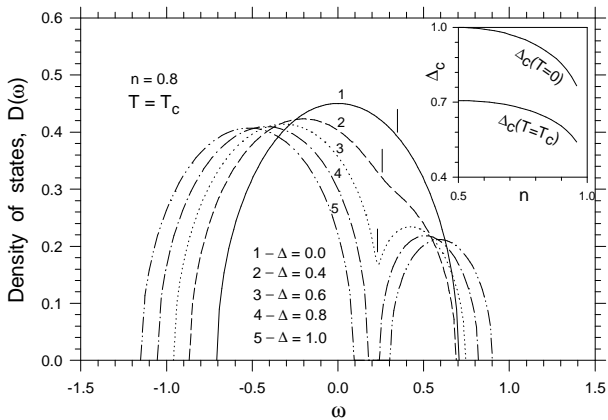}
        }
\end{center}
        \caption{Density of states as a function of frequency for different
disorder strengths $\Delta$ (in the paramagnetic phase). Vertical lines indicate
the location of the chemical potential. Note that the density of states
is plotted on an absolute energy scale, so the chemical potential is
not shifted to lie at $\omega=0$. The inset displays the dependence of
the critical values of disorder in the paramagnetic
[$\Delta(T=T_c)\equiv\Delta_c^P$] and ferromagnetic
[$\Delta_{c\uparrow}(T=0)\equiv\Delta_c^F$] phases as a function of the
electron filling $n$.
       }
\label{fig:1}
\end{figure}

Figure 1 shows the influence of disorder on the conduction-electron
density of states in the paramagnetic phase at $n=0.8$. When
the strength of the disorder is small, the band is only slightly
distorted from the semicircular shape.  As the disorder increases, a
pseudogap first appears and then a true gap develops when the disorder
is larger than a critical value.  Because the double-exchange couples
ferromagnetism to the mobility of the itinerant electrons, the critical
value of the disorder depends on the spin polarization of the system and
on the total electron density.  This is depicted in the inset to Figure
1, where the critical value of the disorder, required for the metal-insulator
transition, is plotted as a function of the total electron density.  Two curves
are shown: (i) the critical disorder strength ($\Delta_c^P$)
needed for the transition
in the high-temperature paramagnetic phase $T>T_c$, where there is no
spin polarization (bottom curve); and (ii) the critical disorder strength
($\Delta_c^F$)
in the fully polarized ferromagnetic phase at $T=0$ (top curve).  Since the
ferromagnetic order makes the bandwidth larger, the critical value of disorder
needed for the MIT increases as one enters the ferromagnetic phase.  If the
disorder is large enough, it is always an insulator (even in the ferromagnetic
phase)---but there is a regime, where the system can be an insulator in the
paramagnetic phase, and a metal in the ferromagnetic phase (at least it is
metallic for the majority [spin-up] electrons, it would be insulating for the
minority [spin-down]
electrons).  This is the regime that is relevant for the CMR materials.

In the limit where $\Delta\to\infty$, the band is always split, and the upper
subband is pushed to infinite energy and can be neglected. From Eq.~(\ref{e25})
we have
\begin{equation}
        {\rm Im} a(\omega-\mu)=\frac{1}{2}\sqrt{2t^{*2}n-\omega^2},
	\qquad
	\omega^2\leq 2t^{*2}n,
	\label{e27}
\end{equation}
for $H=0$.  The magnetization vanishes, because the internal molecular
field $\lambda_F$ vanishes when $\Delta\to\infty$ as seen from Eq.~(\ref{e11}).
Hence, there is no ferromagnetic order, and the inverse of the effective
medium satisfies $a_{\uparrow}(\omega)=a_{\downarrow}(\omega)=a(\omega)$. Since
\begin{equation}
        \frac{2}{\pi t^{*2}}\int d\omega{\rm Im} a(\omega-\mu)=n,
	\label{e28}
\end{equation}
for this case, the lower band is completely occupied and the chemical
potential lies in the gap for all $n$.

For finite $\Delta$, the same conclusion is obtained for the high-temperature
paramagnetic phase.  This follows directly from the numerical calculations,
but can be understood from the fact that the total spectral weight in the
lower band doesn't change until the gap closes.  The ferromagnetic transition
temperature generically increases from zero though.
Figure 1 shows the location of the chemical potential at $T_c$ with
vertical lines for $\Delta=0$, $\Delta=0.4$ and $\Delta=0.6$ (note that
in the last case
the chemical potential lies in the pseudo-gap). For $\Delta=0.8$ and
$\Delta=1.0$ ($\Delta_c=0.6581$ at $n=0.8$), the chemical potential lies in the
gap.

This situation is similar to what takes
place in the static Holstein model, which is solved exactly in infinite
dimensions\cite{r34,r35}. In this model, bipolaron formation
leads to the opening of a gap when the electron-phonon interaction is
strong enough, and
the chemical potential lies in the gap for all $n$. It differs from the
Mott-Hubbard  transition which occurs only at half-filling.

\section{Magnetic properties}
\label{magn}
The magnetism of the binary-disorder double-exchange system
is determined by the uniform (ferromagnetic) magnetization
\begin{equation}
	m=(1-n)m\Bigl(\frac{1}{2}\Delta\Bigr)+nm\Bigl(
	-\frac{1}{2}\Delta\Bigr),
	\label{e29}
\end{equation}
and by the uniform (ferromagnetic) susceptibility
\begin{equation}
	\chi=(1-n)\frac{dm(\frac{1}{2}\Delta)}{dh}+n\frac{dm(-\frac{1}{2}
	\Delta)}{dh}.
	\label{e30}
\end{equation}
The algebraic equations for $dm(\Delta/2)/dh$ and
$dm(-\Delta/2)/{dh}$ are taken at $h=H=m=0$ in Eq.~(\ref{e22}) with the binary
probability distribution for the disorder. The
determinant of this system of equations equals zero at the ferromagnetic
Curie temperature $T_c$.

\begin{figure}[htb]
\begin{center}
        \resizebox{0.45\textwidth}{!}{%
        \includegraphics*{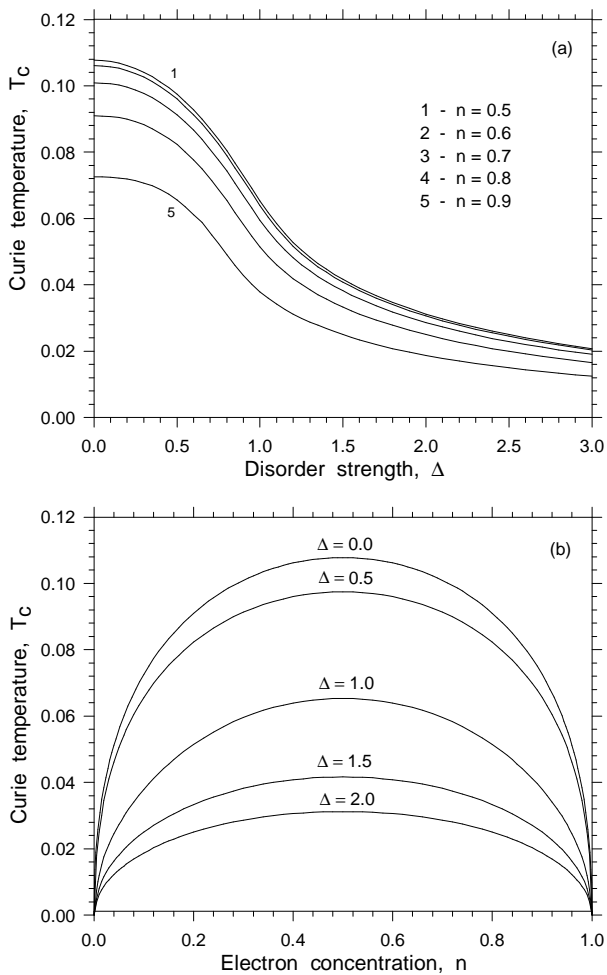}
        }
\end{center}
\caption{Curie temperature as a function of disorder strength $\Delta$ and
electron filling $n$: (a) shows results for constant electron filling,
while (b) shows results for constant disorder strength.
        }
        \label{fig:2}
\end{figure}

Figure 2 shows (a) the disorder dependence of the Curie temperature
for different electron densities and (b)
the electron density dependence of the Curie temperature for different
disorder strengths. The value of $T_c$ for the pure
double-exchange system [the curve $\Delta=0$ in Figure 2(b)] essentially
coincides with that obtained by Furukawa for double exchange with classical
local spins~\cite{r12}. One needs to choose the electronic bandwidth
to be quite narrow ($W$ on the order of $0.5-1$~ev for $T_c$ ranging from
$315-630^o$K at $n\sim 0.5$).
These estimates of $T_c$ for the pure double-exchange system are comparable
with some materials (like LSMO where $T_c\sim 380^o$K), and are much smaller
than those predicted in Ref.\onlinecite{r6}.
Nevertheless, if we take a more reasonable value for the bandwidth
($W\sim 2$~ev), then it is clear
that double exchange alone cannot explain the values of $T_c$ for the
manganites.

Figure 2 shows that disorder suppresses the ferromagnetic transition
temperature. The physics
of this is clear. Carrier motion promotes the double-exchange ferromagnetic
order, whereas disorder reduces the electron motion, and thereby it
reduces the ability of the double-exchange process to produce
ferromagnetism---the net effect is to reduce $T_c$.
The calculated disorder dependence of $T_c$ agrees qualitatively with that found
for the combined double-exchange--Holstein model\cite{r7} and with Narimanov and
Varma's calculation\cite{r22} with a Gaussian disorder probability distribution.

Figure 2 shows also that three disorder regimes can be distinguished for
our model.
The first regime (renormalized double-exchange) is the regime where $T_c$
depends only weakly on disorder and corresponds to the flat regions of the
curves in Figure ~2(a) [$\Delta<0.5$].
We find that $T_c$ is suppressed most strongly at $0.5<\Delta<
1.0$ [see Figure 2(a)]; i.e. in the vicinity of the critical values of disorder
where  a gap in the density of states is created. We call this regime the
transition regime, where the ferromagnetism process is changing from a
renormalized double-exchange process to a renormalized strong-coupling
process.  In this range of
$\Delta$, two crossovers occur: (i) the metallic conductivity is replaced by a
thermally activated (insulating) conductivity, and (ii) double-exchange
ferromagnetism is replaced by a
ferromagnetism that is caused by virtual electron transfers from the
filled lower band to the empty upper subband (and vice versa). Indeed, the
double-exchange mechanism for the ferromagnetic ordering of the local
spins is caused by real electron transfers, which are possible only at small
disorder. For strong disorder $(\Delta > 1$),
when the itinerant electrons are localized,
the ferromagnetic ordering is caused by virtual transfers
to neighboring sites (that have an additional local potential $\Delta$)
and back. (All sites without this
additional potential are filled so hopping onto those sites is impossible.)
Hence, the Curie temperature is inversely proportional to $\Delta$ at
strong disorder [the $1/\Delta$ behavior is clearly seen in Figure~2(a)
for $\Delta>1.5$].

Usually, virtual electron transfers lead to a Heisenberg type of ferromagnetic
ordering, i.e. the magnetic energy contains a $\cos{\Theta}$ dependence, not
the $\cos{\Theta/2}$ dependence of double exchange. Therefore, the
crossover from double-exchange ferromagnetism
to disorder-induced ferromagnetism (with a Heisenberg type of magnetic energy)
must be seen in the changing of the  behavior of magnetic quantities (for
example, paramagnetic susceptibility and magnetization)
as the disorder strength increases.

In particular, an increase of the disorder strength changes the
temperature dependence
of the uniform susceptibility. Indeed,  the pure double-exchange
system ($\Delta=0$) reveals Curie-type behavior for $\chi$
at high temperatures $(\chi\sim1/T$), i.e. the Curie temperature
is equal to zero. As the temperature is decreased,
the curvature of the inverse susceptibility
$\chi^{-1}$ changes, so that it
intersects the temperature axis at a finite temperature, yielding
a nonzero value for $T_c$ and a Curie-Weiss law for $\chi$. This feature of
double exchange was noted in the pioneering work by Anderson and
Hasegawa~\cite{r3} (see also Ref.~\onlinecite{r28}). (It should be noted
that this aspect of double exchange is still not completely understood.
See, for example, Ref.~\onlinecite{r36}
where a high-temperature expansion is employed).

Numerical calculation of the uniform susceptibility, Eq.~(\ref{e30}),
at strong disorder shows that $\chi$ obeys a Curie-Weiss law
with a nonzero Curie temperature in this case. Thus, the
crossover to a disordered-induced mechanism of ferromagnetic ordering
leads to a change in the temperature dependence of $\chi$.


\begin{figure}[htb]
\begin{center}
        \resizebox{0.45\textwidth}{!}{%
        \includegraphics*{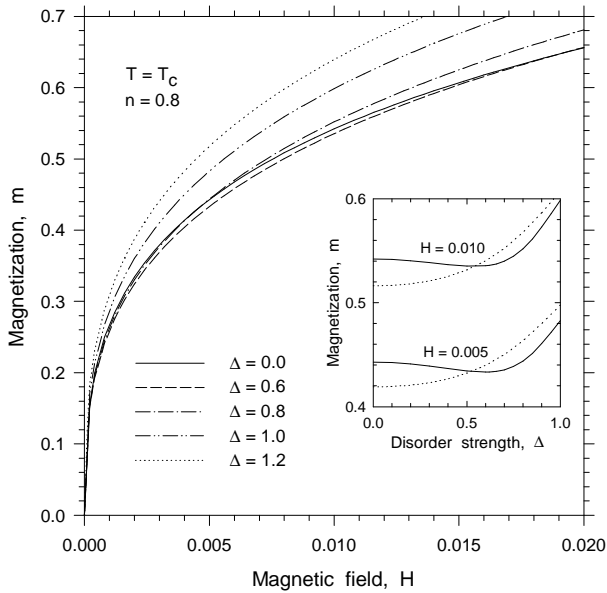}
        }
\end{center}
\caption{Magnetization of the local-spins $m$ as a function of the
         external magnetic field $H$ for different disorder strengths $\Delta$.
  	 The inset shows the disorder dependence of $m$ at $H=0.005$
	 and $H=0.01$ for the double-exchange model (solid lines) and for
	 the Ising model in the mean-field approximation (dotted lines).
        }
        \label{fig:3}
\end{figure}

In Figure 3 we plot the magnetization $m$ as a function of external
magnetic field $H$ at $T=T_c$ for a number of different disorder strengths.
The field-induced magnetization at fixed magnetic field depends on disorder
and the inset to Figure 3 shows that the magnetization (solid lines)
initially decreases as the disorder strength increases until one
reaches the region where the conduction band has a well-developed pseudogap
and begins to split ($\Delta_c=0.6581$ for $n=0.8$). As the disorder increases
further, the magnetization starts to increase and it can be described
by the following mean-field equation
\begin{equation}
	m=\tanh{\beta\Bigl(mT_c+\frac{1}{2}H\Bigr)},
	\label{e31}
\end{equation}
for the magnetization of an Ising model (or a Heisenberg model with $S=1/2$).
Dotted lines in the inset show the field-induced magnetization evaluated
from (\ref{e31}) where $T_c$ is equal to the Curie temperature of the
disordered double-exchange model at a given $\Delta$. Hence, the
disorder dependence of the magnetization at fixed magnetic field
also shows the crossover from double-exchange-induced to
disorder-induced ferromagnetism at strong disorder.

\begin{figure}[htb]
\begin{center}
        \resizebox{0.45\textwidth}{!}{%
        \includegraphics*{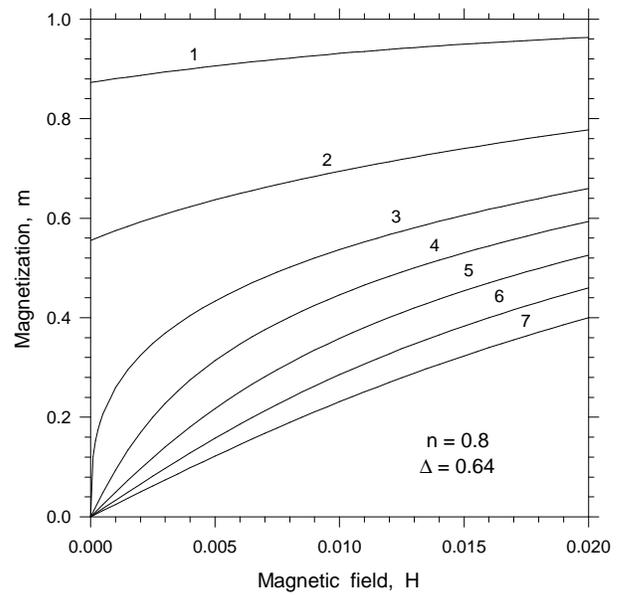}
        }
\end{center}
\caption{Magnetization of the local-spins, $m$, as a function of
        the external magnetic field $H$ for different relative temperatures
	$\tau=T/T_c$: 1 - $\tau=0.4$, 2 - $\tau=0.8$, 3 - $\tau=1.0$,
	4 - $\tau=1.1$, 5 - $\tau=1.2$, 6 - $\tau=1.3$, 7 - $\tau=1.4$.
        }
        \label{fig:4}
\end{figure}

From Figure 3 we see that the most favorable conditions for a large
field-induced magnetization are strong disorder and strong magnetic fields.
Figure 4 shows the field dependence of $m$ for different relative temperatures
$\tau=T/T_c$ and fixed disorder. The largest absolute value for $m$
is observed at small $\tau$ ($\tau=0.4$) but the relative increase of
$m$ (when $H$ is also increased) is weak. The strongest growth of $m$
at weak fields is seen near the Curie temperature. At $T=T_c$
($\tau=1.0$) this growth is maximal, but even at $\tau=1.4$ the $H$ dependence
of $m$ is linear. Note that such behavior of the field-induced
magnetization is in good agreement with the
experimental data on the manganites~\cite{r14}.

Our analysis of the magnetic properties has shown the emergence of three
different regimes: (i) a renormalized double-exchange regime (or weak-disorder
regime); (ii) a strong-coupling physics regime (or strong-disorder regime);
and (iii) a transition regime where the properties crossover from metal
to insulator and from double-exchange ferromagnetic
ordering to strong-coupling-induced ferromagnetism.
This classification scheme will be more sharply defined as we examine the
transport properties in Section V.

\section{Transport properties}
\label{transport}
In this section, we consider the electrical and thermal
transport properties of the double-exchange system in zero external magnetic
field. The transport is determined from the following set of
equations\cite{r37}. The electrical conductivity $\sigma=1/\rho$ is
\begin{equation}
	\sigma=L_{11};
	\label{e32}
\end{equation}
the thermopower is
\begin{equation}
	S=-\frac{k_B}{|e|}\frac{1}{T}\frac{L_{12}}{L_{11}},
	\qquad
	\frac{k_{B}}{|e|}\simeq 86\mu\mbox{VK$^{-1}$};
	\label{e33}
\end{equation}
and the thermal conductivity (of the electronic system) satisfies
\begin{equation}
	\kappa
=\biggl(\frac{k_B}{e}\biggr)^2\frac{1}{T}\Bigl\{L_{22}-\frac{L_{12}^2}
	{L_{11}}\Bigr\};
	\label{e34}
\end{equation}
where the transport coefficients $L_{ij}$ are defined by
\begin{eqnarray}
	L_{ij}&=&\pi\sigma_0\sum_{\sigma}\int\limits_{-\infty}^{\infty}
	d\varepsilon v^2(\varepsilon)D^0(\varepsilon)\cr
        &\int\limits_{-\infty}^{\infty}&
        d\omega\biggl(-\frac{df(\omega)}{d\omega}\biggr)
	(\omega-\mu)^{i+j-2}A^2_{\sigma}(\varepsilon,\omega-\mu).
	\label{e35}
\end{eqnarray}
Here the spectral function is given by
\begin{equation}
	A_{\sigma}(\varepsilon,\omega)=-\frac{1}{\pi}\mbox{Im}
	{\cal G}_{\sigma}(\varepsilon,\omega),
	\label{e36}
\end{equation}
$\sigma_0$ is the unit of conductivity, ${\cal G}_{\sigma}(\varepsilon,\omega)$
is given in Eq.~(\ref{e14}) and
$v(\varepsilon)$ is the current vertex which is equal to
\begin{equation}
	\sqrt{(4t^{*2}-\varepsilon^2)/3},
	\label{e37}
\end{equation}
for the Bethe lattice\cite{r38,r39} with $z\to\infty$.

Figure 5 shows the temperature dependence of the resistivity
for different electron fillings in the three different
disorder regimes. (In this and the
following Figures, the resistivity is plotted in units of $\rho_0=1/\sigma_0$).
In Figure 5(a), the weak-disorder regime, the
transport properties are determined by the double-exchange mechanism.
From Eqs.~(\ref{e17}) and (\ref{e25}), one
obtains the following expression for the electronic self-energy
        $$
        \Sigma_{\sigma}(\omega+i\delta)=-\frac{1}{2}\biggl(\frac{1-\sigma m}
        {1+\sigma m}\biggr)
        $$
\begin{equation}
        \times\Bigl\{\omega+\mu\pm\sqrt{(\omega+\mu)^2-
        2t^{*2}(1+\sigma m)}\Bigr\},
	\label{e38}
\end{equation}
at $\Delta=0$. Therefore,
        $$
        {\rm Im}\Sigma_{\sigma}(\omega+i\delta)=-\frac{1}{2}
        \biggl(\frac{1-\sigma m}{1+\sigma m}\biggr)
        $$
\begin{equation}
        \times\sqrt{2t^{*2}(1+\sigma m)-(\omega+\mu)^2},
	\label{e39}
\end{equation}
for $(\omega+\mu)^2\leq2t^{*2}(1+\sigma m)$ and $\mbox{Im}\Sigma_{\sigma}
(\omega+i\delta)=0$ otherwise.

Note that the factor
\begin{equation}
	\frac{1-\sigma m}{1+\sigma m},
\end{equation}
in Eq.~(\ref{e39}) is contained in the expression for the self-energy in
Ref.\onlinecite{r40} for the pure double-exchange system with classical local
spins and a Lorentzian density of states (we use a semicircular density of
states here). This factor plays an essential role
in the low-temperature behavior of the resistivity [Figure 5(a)]. It provides
a decrease in the resistivity when the magnetization increases. Indeed,
in the limit $m\to1$ or $T\to0$, we have
\begin{equation}
	A_{\uparrow}(\varepsilon,\omega)=\delta(\omega+\mu-\varepsilon),
	\qquad
	A_{\downarrow}(\varepsilon,\omega)=0,
	\label{e40}
\end{equation}
and the conduction-electron subsystem becomes a free-electron gas of
spin-up electrons with a conductivity that is a delta function at
zero frequency (whose strength is the Drude weight).

On the other hand, in the high-temperature paramagnetic phase, $m=0$, and the
correlated bandwidth is narrowed
by a factor of $\sqrt{2}$ due to paramagnetic spin disorder. In this case,
the expression for the self-energy exactly coincides with the one obtained
in Ref.~\onlinecite{r7} for the pure double-exchange system. Since the
density of states is independent of temperature here, all of the temperature
dependence of the resistivity arises from the Fermi factor
$[-df(\omega)/d\omega]$ [see Eq.~(\ref{e35})].
If one makes the assumption that the chemical potential is also
temperature-independent, so that the derivative of the Fermi factor
depends only weakly on temperature, then one would
conclude\cite{r12,r22} that the resistivity in
the paramagnetic phase (due to double exchange only) is essentially
a constant. However,
more accurate calculations, that take into account the temperature dependence
of the chemical potential, show that the pure double-exchange system becomes
a bad-metal with $d\rho/dT>0$
[see, for example, Ref.~\onlinecite{r7} where the bad-metal behavior
was shown in a double-exchange system with weak
electron-phonon interaction in their Figure 6(a)].

The sharp decrease of the resistivity at $T<T_c$ for all $n$ is caused by the
rapid increase of the magnetization $m$.
This discontinuity in the slope $d\rho/dT$ at $T=T_c$ is a consequence of
the dynamical mean-field approach. Incorporation of spatial spin
fluctuations\cite{r27} will smooth out the temperature dependence of the
resistivity in the vicinity of $T_c$, but this is beyond dynamical mean
field theory.

\begin{figure}[b]
\begin{center}
        \resizebox{0.45\textwidth}{!}{%
        \includegraphics*{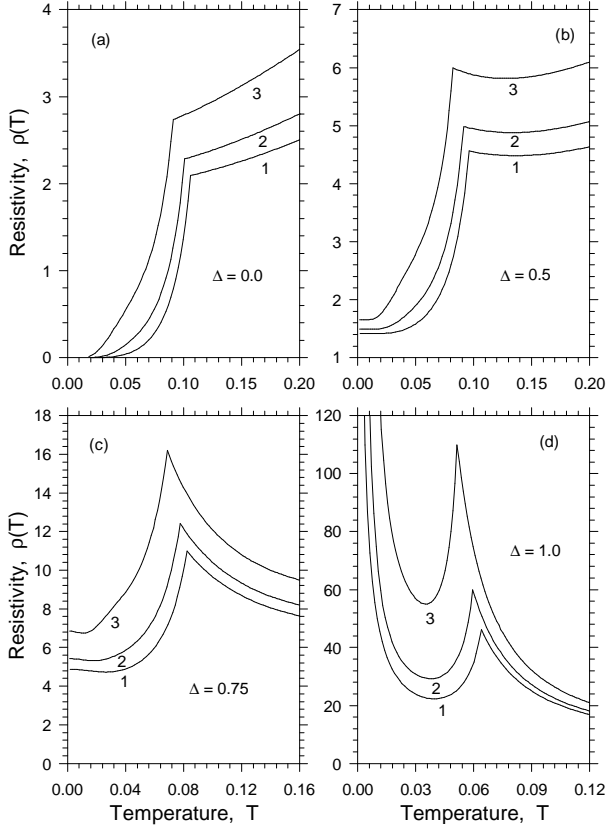}
        }
\end{center}
\caption{Temperature dependence of the resistivity $\rho$ for different
         disorder strengths $\Delta$: (a) $\Delta=0$, (b) $\Delta=0.5$,
         (c) $\Delta=0.75$ and (d) $\Delta=1.0$. The curves labeled by 1
         correspond to an electron filling of $n=0.6$; 2 denotes $n=0.7$;
         and 3 denotes $n=0.8$.  }
        \label{fig:5}
\end{figure}

As disorder is added to the system, the properties are initially changed
little.  When the disorder becomes large enough $\Delta\approx 0.4$, then
we start to feel a more direct influence of the disorder as it produces a
pseudogap in the density of states and we enter the transition regime
of moderate disorder. In this regime, the slope of the resistivity can become
negative $d\rho/dT$  above $T_c$ as shown in
Figures 5(b) and 5(c). The value $\Delta\simeq0.4$ is the boundary value that
separates the weak-disorder and the moderate-disorder regimes for $n=0.8$.
As the disorder is increased further, the
metallic conductivity is gradually replaced by a thermally activated
conductivity (this starts at $\Delta\approx \Delta_c^P$). As one enters the
strong-disorder regime, there is a MIT at $T_c$. Here the system displays
insulating behavior everywhere, except just below the Curie point, where
the rapid increase in the magnetization can cause the resistance to drop
over a small temperature range before it turns around and increases again.
This occurs in the transition from the paramagnetic insulator to the
ferromagnetic insulator because the charge gap in the ferromagnetic
insulator is smaller than the charge gap in the paramagnetic insulator.
Those sharp cusps seen in Figure 5(d) will generically be smoothed out
by spatial fluctuations.

\begin{figure}[htb]
\begin{center}
        \resizebox{0.45\textwidth}{!}{%
        \includegraphics*{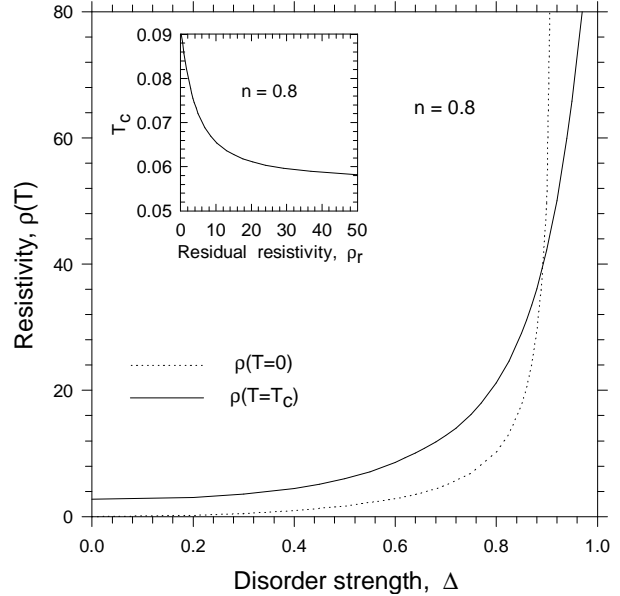}
        }
\end{center}
\caption{Resistivity at $T=0$ (dotted line) and at $T=T_c$
(solid line) versus
disorder strength.  Inset is a plot of the Curie temperature $T_c$
versus the residual resistivity $\rho_r=\rho(T=0)$.  Note how the two
resistivity curves cross at a critical value of disorder, which determines the
beginning of the transition region from the moderate to strong-disorder
regimes.}
\label{fig:6}
\end{figure}

We can analyze the transition from moderate to strong disorder more
quantitatively.  We focus on two characteristic resistivities: the
resistivity at $T_c$ and the resistivity at $T=0$ (residual resistivity).
For weak disorder
the residual resistivity is much smaller than the resistivity
at $T_c$.  When we reach the moderate disorder regime, the residual
resistivity starts to increase rapidly, eventually overtaking the
resistivity at $T_c$ in the strong-disorder regime.  We denote the
boundary between the moderate and strong-disorder regimes by the value
of disorder where $\rho(T=0)=\rho(T=T_c)$.  This occurs at
$\Delta\simeq0.892$ for $n=0.8$ as shown in Figure 6. This transition occurs
when the density
of states has developed a strong pseudogap, but has not yet become an insulator
($\Delta_c^F=0.931$ for $n=0.8$).
The range of disorder between these two limits,
($0.892<\Delta<\Delta_c^F=0.931$  for $n=0.8$) is the transitional
region between the
moderate-disorder and the strong-disorder regimes.

This analysis is in qualitative agreement with the calculations that use
a Lorentzian density of states\cite{r23} and especially with
calculations\cite{r18} that show a divergence of the $T=0$
resistivity at a critical disorder strength.  Note that
Anderson localization is the cause of the MIT in Ref.~\onlinecite{r18}.

\begin{figure}[htb]
\begin{center}
        \resizebox{0.45\textwidth}{!}{%
        \includegraphics*{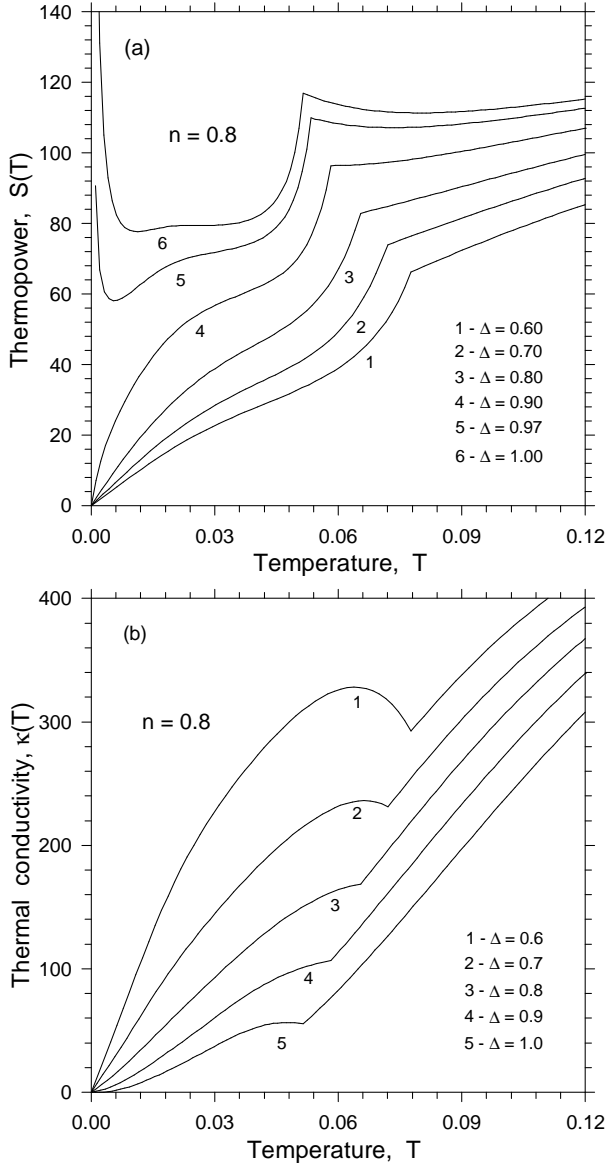}
        }
\end{center}
\caption{Temperature dependence of (a) the thermopower and (b) the
electronic thermal conductivity for different values of disorder.}
\label{fig:7}
\end{figure}

The inset to Figure 6 shows the Curie temperature
as a function of the residual resistivity
$\rho(T=0)$. This plot was obtained by combining the $\Delta$ dependences
of $\rho(T=0)$ and $T_c$ (see Figure 2). It is clearly seen that the
suppression of $T_c$ due to disorder is accompanied by the increase of the
residual resistivity. We find that our functional dependence of $T_c$ on
$\rho(T=0)$ is smoother than that found in Ref.~\onlinecite{r18} and agrees better
with experiment~\cite{r18,r41}.

Now we examine the thermal properties of our system including the thermopower
$S(T)$ and the electronic thermal conductivity $\kappa(T)$ which are
shown in Figure 7. The behavior of these two quantities are quite
different from each other---while the thermal conductivity always vanishes
as $T\to 0$, the thermopower will either vanish or diverge depending on
whether the system is metallic or insulating as $T\to 0$.  The thermal
conductivity behaves as expected with a sharp increase at $T_c$ due
to the opening of conduction channels as the magnetization grows, and
a linear decrease to zero at low temperatures.  Note that the thermal
conductivity vanishes at low temperatures even in the metal, because the
heat current vanishes.

The thermopower behaves quite a  bit differently.  It too shows a strong
effect at $T_c$ (in this case a sharp decrease), but the low temperature
behavior is most interesting.  For weak disorder (metallic phases), the
thermopower vanishes as $T\to 0$, but once a gap opens in the density of states,
the thermopower diverges as $T\to 0$ since the chemical potential lies in the
gap. Peltier's coefficient, $P=TS(T)$, is approximately
given by a straight line for strong disorder ($\Delta=1.0$) and low temperature
(where the magnetization has saturated
$m\simeq1$), i.e. $P=\alpha+\gamma T,$ $\alpha\simeq
0.17\mu$V. Hence, the thermopower $S(T)$ can be approximately represented by
\begin{equation}
	S(T)\simeq\frac{\alpha}{T}+\gamma,
	\label{e41}
\end{equation}
for low temperature and strong disorder. This relation is typical
of what is seen in intrinsic semiconductors~\cite{Mott}
Note that the thermopower behaves similarly for electronic systems that undergo
Anderson localization\cite{r42}:
if the chemical potential is in the region of localized states, then
$S(T)$ also diverges as $T\to0$.

The slope of the thermopower $dS/dT$ has a discontinuity at $T_c$, but $S(T)$
does not change sign in our model.
The prediction\cite{r43} that $S(T)$ alter its sign at $T=T_c$ in
double-exchange systems was founded on two principles:
(i) the itinerant-electron subsystem is a Fermi-liquid and (ii) the
derivative of the chemical potential
$d\mu/dT$ changes sign at $T=T_c$. While we find that the derivative $d\mu/dT$
does change sign at $T=T_c$ in agreement with others\cite{r12,r28}, the
itinerant-electron
subsystem is not a Fermi-liquid in our model, because the imaginary part of
the self-energy does not vanish as $T\to 0$ at the Fermi-energy. Hence, the
reasoning that led to the prediction of a sign change cannot be applied here.

The high-temperature behavior of the thermopower for strong disorder is similar
to that of $S(T)$ in a small-polaron model\cite{r44}
where the conductivity is thermally activated and $S(T)\sim\ln [{c/(1-c)}]$;
with $c$ the small-polaron concentration and $c\leq1$.
Indeed, the thermopower has a weak temperature dependence
for strong disorder [see Figure 8(a)] in the paramagnetic phase and
$T\gg T_c$. Furthermore, for fixed temperature, we find that  $S(T)$
decreases when the electron filling is decreased to $n=0.5$.
At $n=0.5$ the thermopower is equal to zero at all temperatures as in the
small-polaron theory.
Below $n=0.5$, we find that the thermopower changes sign.  Hence we only
find a sign change of the thermopower when the electron filling is varied.

\section{Magnetoresistance}
\label{CMR}
The most interesting property of the manganites is the fact that the resistance
changes so dramatically in a magnetic field.  This makes them useful as
possible magnetic field sensors for the magnetic storage community.  We find
similar magnetoresistance effects in our model, especially when we are close
to the Curie point.  The origin of the magnetoresistance lies in the
sensitivity of the resistivity to the magnetization, and the ease with which
$m$ can be tuned by a magnetic field.  Typically, the field increases $m$,
which then reduces the resistivity.  Experimentally\cite{r14}, there is
a strong correlation between the field-induced changes in
$\rho$ and $m$.

In order to calculate the magnetoresistance, we must first solve for the
conductivity in the presence of an external magnetic field.  If we perform
a simple shift in the definition of the inverse effective medium
$a_{n\sigma}\to a_{n\sigma}+H\sigma/2$ and replace the integration variable
in Eq.~(\ref{e25}) by $\omega\to\omega+H\sigma/2$, then the only modification
is that the derivative of the Fermi factor $[-df(\omega)/d\omega]$ now has
an explicit $H$ dependence and the spectral function $A_{\sigma}$
in Eq.~(\ref{e27}) depends on $H$ through the magnetization $m$.
The direct dependence of $\rho$ on $H$ via the factor
$[-df(\omega)/d\omega]$ always yields a small positive magnetoresistance that
is caused by the Zeeman interaction with the external magnetic field; it
can be neglected in weak fields.

\begin{figure}[htb]
\begin{center}
        \resizebox{0.45\textwidth}{!}{%
        \includegraphics*{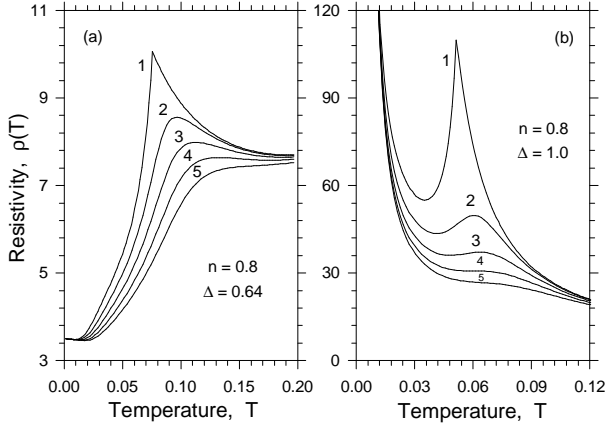}
        }
\end{center}
\caption{Temperature dependence of the resistivity $\rho(T)$ for different
magnetic fields $H$: 1 - $H=0.0$, 2 - $H=0.005$, 3 - $H=0.01$, 4 - $H=0.015$,
        5 - $H=0.02$.}
\label{fig:8}
\end{figure}

Figure 8 shows the temperature dependence of the resistivity for different
magnetic fields and two disorder regimes: (a) moderate-disorder and (b)
strong-disorder. The field-induced modifications to the magnetization
suppress the resistivity. This effect is strongest
in the vicinity of the Curie point, because the spin susceptibility is
large there; hence the field changes the magnetization most strongly there.
The peak position in $\rho(T)$ shifts to higher
temperature with increasing $H$ and there is a critical value of $H$
above of which the MIT disappears (i.e., $d\rho/dT>0$ for all
temperatures). This picture qualitatively agrees with the experimental data
on manganites~\cite{r1,r14}.

\begin{figure}[htb]
\begin{center}
        \resizebox{0.45\textwidth}{!}{%
        \includegraphics*{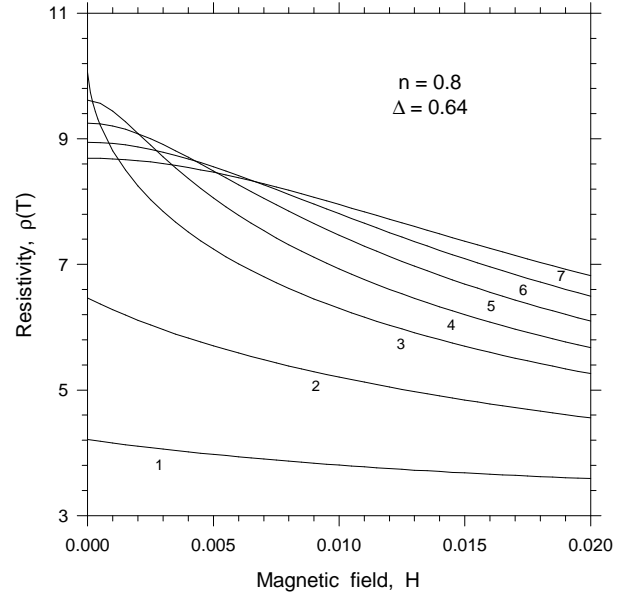}
        }
\end{center}
\caption{Resistivity $\rho(T)$ as a function of external magnetic field $H$
         for different relative temperatures $\tau=T/T_c$: 1- $\tau=0.4$,
         2 - $\tau=0.8$, 3 - $\tau=1.0$, 4 - $\tau=1.1$, 5 - $\tau=1.2$,
         6 - $\tau=1.3$ and 7 - $\tau=1.4$.}
\label{fig:9}
\end{figure}

The magnetic-field dependence of the resistivity
at different relative temperatures $\tau=T/T_c$
(Figure 9) also indicates the close correlation between the
field-induced  changes in $\rho$ and $m$. Indeed, the comparison of Figures
4 and 9 shows that the magnitude of resistivity correlates with the change
in the magnetization: the suppression of the resistivity is large around $T_c$
where the largest growth of $m$ is seen (see Figure 4) but is
weak far below and above $T_c$. Our calculation of the $H$ dependence
of $\rho$ is in good qualitative agreement with experiment (see Figure 7
in Ref.~\onlinecite{r14}).

We define the magnitude of the relative magnetoresistance as
\begin{equation}
        \frac{\delta\rho}{\rho}=\frac{\rho(T,0)-\rho(T,H)}{\rho(T,0)}.
        \label{e42}
\end{equation}
In this definition, $\delta\rho/\rho$ is positive and cannot exceed 100\%.
Figure 10 shows the magnetoresistance as (a) a function of $H$ and (b) as a
function of $m^2$ for different disorder strengths at $T=T_c$.
It is seen from the comparison of Figure 3 and Figure 10(a) that the
field-dependence of the magnetization is
directly reflected in the field-dependence of the magnetoresistance.
The $\Delta$ dependence of $\delta\rho/\rho$ at fixed $H$
is shown in the inset to the Figure 10(a). The comparison of this inset with the
inset to  Figure 3 shows that disorder dependences of $\delta\rho/\rho$ and
$m$ at fixed $H$ are also similar. In the weak-disorder regime,
$\delta\rho/\rho$ is slightly decreased. Although the double-exchange mechanism
of ferromagnetic ordering becomes weaker in this regime, it is still
active.  In the
moderate-disorder regime, strong-coupling-induced ferromagnetism (discussed in
Section IV) replaces the double exchange,
 and the magnetization begins to increase
with increasing disorder. This leads to an increase of the magnetoresistance
for both the moderate-disorder and strong-disorder
regimes. At $\Delta=1.0$, $\delta\rho/\rho$ can attain values near 70\% in a
weak field of $H=0.01$.

\begin{figure}[htb]
\begin{center}
        \resizebox{0.45\textwidth}{!}{%
        \includegraphics*{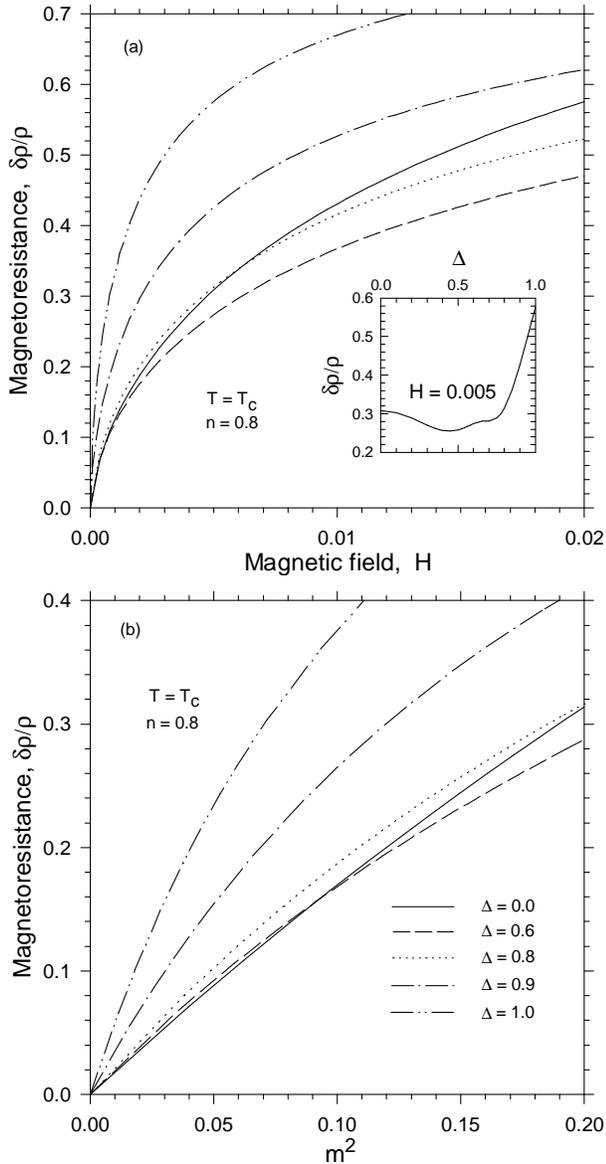}
        }
\end{center}
\caption{(a) Magnetoresistance $\delta\rho/\rho$ as a function of
the external magnetic field $H$ and (b) magnetoresistance as a
function of $m^2$ for different
disorder strengths. The inset shows the disorder-dependence of
$\delta\rho/\rho$ for $H=0.005$. The line styles in (a) are
the same as in (b).}
\label{fig:10}
\end{figure}

Figure 10(b) shows that the magnetoresistance can be expressed by a scaling
law
\begin{equation}
\frac{\delta\rho}{\rho}=Cm^2,
	\label{e43}
\end{equation}
where the scaling constant $C$ is independent of $T$ only for the pure
double-exchange system ($\Delta=0$). The relation
(\ref{e43}) is approximately satisfied for finite disorder
$0\leq\Delta\leq1.0$ (at least for $m^2<0.01$), but
the coefficient $C$ is rather high, about 4, for $\Delta=1.0$
($n=0.8$). versus its value of 1.9 for weak-to-moderate-disorder
$(0<\Delta<0.6)$.  Note that Furukawa's calculation~\cite{r45}
performed for the pure-double exchange system with classical local
spins gives the value of 4 for $C$ (when
$n$ is also equal to 0.8). The difference
between our estimate of $C$ and Furukawa's arises from the different
density of states.  Note that Kubo
and Ohata~\cite{r5} obtain $C=1$ for the quantum double-exchange system.
The scaling constant $C$ also depends on band filling. $C$ decreases
with decreasing electron filling. In particular, $C=1.7$ $(\Delta=0)$ and
$C=2.5~(\Delta=1.0)$ for $n=0.67$.
Thus, we can conclude that the band filling near $n=0.8$ and relatively strong
disorder ($\Delta>0.9$) are the most favorable conditions for a large
magnetoresistance.

\section{Conclusion}
\label{Conclusion}
We have considered the influence of diagonal disorder on a
simple double-exchange model with local Ising spins by employing dynamical
mean-field theory. We choose a binary probability distribution for the disorder
which greatly simplified the analysis and allowed us to examine directly
the MIT.
The manganites are too complicated a system to be described completely by
this simple model. Nevertheless, we still arrive at some useful
conclusions: (i) double exchange alone
cannot explain the metal-insulator transition in manganites
(in the best case, it can be applicable to LSMO at $x\simeq0.3$
which has a relatively high Curie temperature and displays bad-metal behavior
in the paramagnetic phase); (ii) double exchange plus disorder
cannot explain the temperature dependence of the thermopower $S(T)$ because it
does not yield a change of sign in the paramagnetic phase. An explanation
of the large peak in $S(T)$ (in the ferromagnetic phase) is beyond dynamical
mean field theory, because it does not include magnon drag\cite{r46};
(iii) the effect of diagonal disorder (induced by chemical substitution)
is important, and can influence the properties of the material if the disorder
strength is large enough to be close to the MIT;
and (iv) we identified three disorder regimes which display different
characteristic behavior.  The weak-disorder regime has little effect on the
system and just renormalizes the double-exchange mechanism of ferromagnetic
ordering. The moderate-disorder regime, corresponds to the transition region
between weak and strong disorder.  The interacting density of states develops
a pseudogap that promotes behavior similar to thermal activation.
The double-exchange mechanism is gradually
replaced by a strong-coupling ferromagnetism (see Section IV) which has a
mean-field-like magnetism. In this regime, the Curie temperature
is sharply decreased and the temperature dependence of the resistivity
reveals a MIT at $T=T_c$. The strong-disorder
regime, is characterized by a gap in the interacting density of states
at high temperature and the residual resistivity exceeds the resistivity
at $T=T_c$.
Ferromagnetic ordering of local spins (with small values of $T_c
\sim1/\Delta$) occurs due to strong-coupling physics. This regime
is most favorable to obtain a large magnetoresistance.

\acknowledgments

B.M.L. acknowledges support from the Project
for supporting of Scientific Schools, N 00-15-96544
and J.K.F. acknowledges support from the National Science Foundation under
grant number DMR-9973225.
We acknowledge useful discussions with J. Byers and D. Edwards.

\end{document}